\documentclass[conference, onecolumn]{IEEEtran}
\usepackage{cite}
\usepackage{amsmath,amssymb,amsfonts}
\usepackage{algorithmicx}
\usepackage{graphicx}
\usepackage{textcomp}
\usepackage{xcolor}
\usepackage{hyperref}
\usepackage{booktabs}
\usepackage{multirow}
\usepackage[utf8]{inputenc}
\usepackage{algorithm}
\usepackage{caption}
\usepackage{graphicx}
\usepackage{caption}
\usepackage{subcaption}
\usepackage{lipsum}
\usepackage{algpseudocode}

\def\BibTeX{{\rm B\kern-.05em{\sc i\kern-.025em b}\kern-.08em
    T\kern-.1667em\lower.7ex\hbox{E}\kern-.125emX}}
    
\begin{document}
\title{MLP-Enhanced Nonnegative Tensor RESCAL Decomposition for Dynamic Community Detection}

\author{
\IEEEauthorblockN{Chaojun Li}
\IEEEauthorblockA{College of Computer and Information Science\\
Southwest University\\
Chongqing, China\\
2938643314@qq.com}
\and
\IEEEauthorblockN{Hao Fang*}
\IEEEauthorblockA{College of Computer and Information Science\\
Southwest University\\
Chongqing, China\\
haof59343@gmail.com}
}

\maketitle

\begin{abstract}
Dynamic community detection plays a crucial role in understanding the temporal evolution of community structures in complex networks. Existing methods based on nonnegative tensor RESCAL decomposition typically require the decomposition rank to equal the number of communities, which limits model flexibility. This paper proposes an improved MLP-enhanced nonnegative tensor decomposition model (MLP-NTD) that incorporates a multilayer perceptron (MLP) after RESCAL decomposition for community mapping, thereby decoupling the decomposition rank from the number of communities. The framework optimizes model parameters through a reconstruction loss function, which preserves the ability to capture dynamic community evolution while significantly improving the accuracy and robustness of community partitioning. Experimental results on multiple real-world dynamic network datasets demonstrate that MLP-NTD outperforms state-of-the-art methods in terms of modularity, validating the effectiveness of the proposed approach.
\end{abstract}

\begin{IEEEkeywords}
Dynamic community detection, tensor decomposition, multilayer perceptron, modularity maximization, dynamic networks
\end{IEEEkeywords}

\section{Introduction}
In today's interconnected world, relationships between entities evolve in complex ways, making dynamic community detection an essential tool for understanding evolution patterns in social networks, biological networks, and information systems\cite{li2023dynamic,22chen2021efficient,22fang2024modularity,22hu2025comprehensive,22liao2025novel,22liao2025proximal,22liu2024hp,22luo2023neulft,22wu2024fine,22wu2022pid,22luo2021adjusting,22wang2024dynamically,10946009,10949771,G} . Dynamic community detection aims to identify evolving community structures in dynamic networks, where nodes within communities exhibit dense connectivity while connections between communities remain sparse\cite{11fang2025nmtd,22chen2021efficient,22liao2025novel,22luo2020temporal,22liu2024hp,22wu2021proportional,22chen2025adaptive,22luo2023neulft,22tang2024temporal,22wu2021discovering,22wu2021instance,11002375,H,I}, ultimately revealing key patterns such as community evolution, information diffusion, and functional module migration\cite{liu2020detecting,22liao2025novel,22zhang2025prediction,11025153,11033266,J} .

Traditional dynamic community detection methods primarily rely on non-negative matrix factorization  and its tensor extensions \cite{luo2022symmetric,11wu2023dynamic,11wu2025learning,11chen2025adaptive,11036662,11059286,K}. Among these,tensor RESCAL decomposition has significant theoretical and practical advantages compared to ordinary tensor decomposition\cite{11164288,11131623,11114958,11235568}, it has gained significant attention due to its capability to effectively capture latent community structures across multiple time slices . However, existing methods face a critical limitation: the decomposition rank must equal the number of communities\cite{11wang2025convolution,11wu2024fine,11220950,11216416,11082554,F}, When the number of communities is small, an excessively small designed rank will lead to the loss of global detailed information of nodes, thereby resulting in misjudgment; conversely, an excessively large rank will cause low computational efficiency and poor stability\cite{11082558,11267233,11231130,D,E}.

To address this limitation, this paper proposes incorporating a MLP mapping layer after nonnegative tensor RESCAL decomposition to obtain initial community affiliations, which is subsequently processed by a modularity maximization algorithm to optimize community assignments. The main contributions include:

\begin{itemize}
\item \textbf{Decoupled Community Mapping}: An MLP maps latent representations from RESCAL output to community space, enabling independent setting of decomposition rank and number of communities.
\item \textbf{Reconstruction Loss Optimization}: A loss function based on adjacency matrix reconstruction ensures the model maintains community structure while optimizing mapping relationships.
\end{itemize}

Experimental results demonstrate that MLP-NTD outperforms the original MNTD and other comparative methods across multiple evaluation metrics, particularly in scenarios with dynamically changing community numbers.

\section{Preliminaries}

\subsection{Notations}
Let a set of undirected temporal networks $\mathbf{G} = (G_1, G_2, \ldots, G_T)$ represent a dynamic network, where $T$ denotes the number of time slices, and each time slice $G_t$ contains $N$ nodes. The adjacency matrix for each time slice is denoted as $\mathbf{W}_t \in \mathbb{R}^{N \times N}$, where element $w_{ijt}$ indicates the presence (1) or absence (0) of an edge between nodes $i$ and $j$ at time $t$.

\subsection{Problem Description}
The community indicator matrix sequence is defined as $\mathbf{B} = (B_1, B_2, \ldots, B_T)$, where $B_t \in \mathbb{R}^{N \times K}$ and $K$ represents the number of communities. Element $b_{ijt}$ indicates the probability that node $i$ belongs to community $j$ at time $t$. The final community partition is denoted as $\mathbf{C} = (c_1, c_2, \ldots, c_T)$, where $c_t$ represents the community assignment vector at time $t$\cite{9238448,8840875,A,B,C}.

\begin{figure}[htbp]
  \centering
  \includegraphics[width=\linewidth]{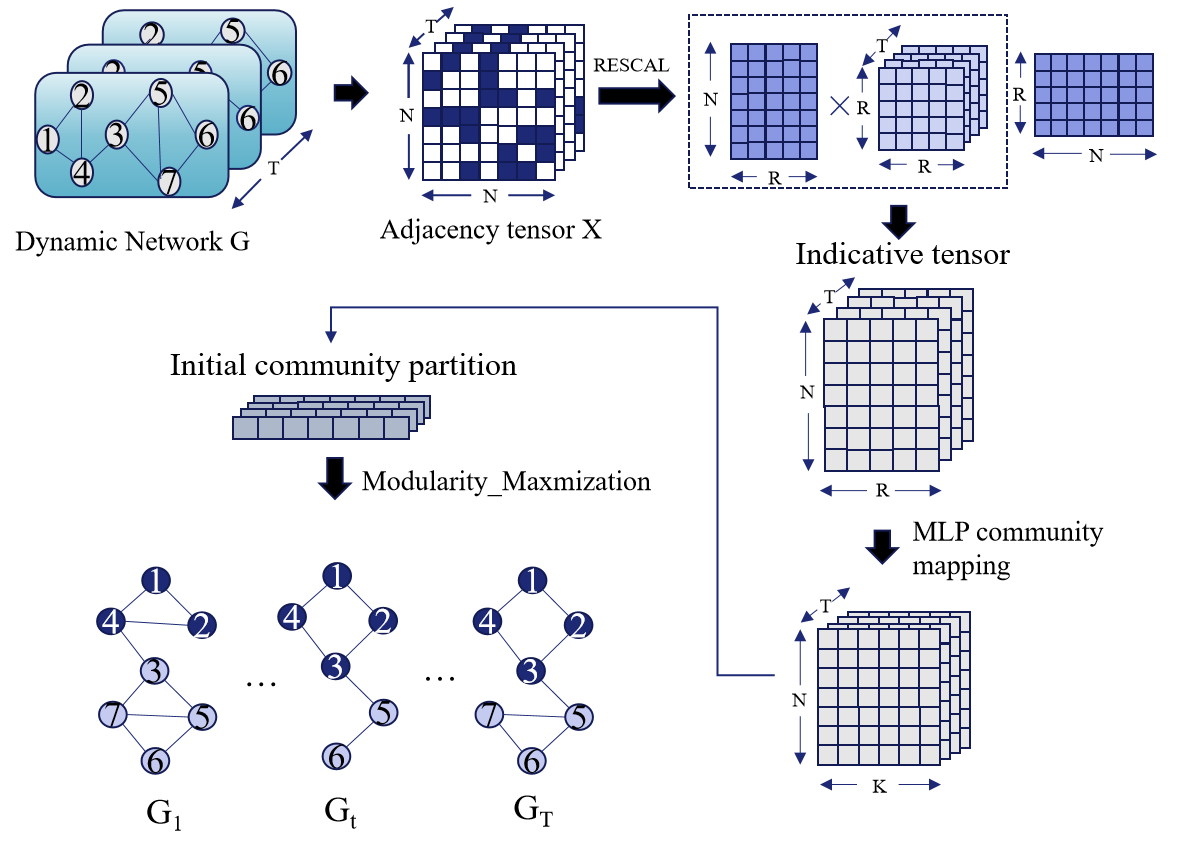}
  \caption{Framework of the MLP-NTD model}
  \label{fig:model_framework}
\end{figure}

\section{The MLP-NTD Model}
This section formulates the optimization objective of the proposed method, presents the nonnegative update rules incorporating MLP training, and analyzes the computational complexity of the algorithm. The comprehensive workflow of the MLP-NTD model is illustrated in Fig. 1.

\subsection{Model Architecture}
The MLP-NTD model consists three core components:

\subsubsection{Nonnegative Tensor RESCAL Decomposition}
The three-order adjacency tensor $\mathbf{X} \in \mathbb{R}^{N \times N \times T}$ is decomposed as:
\begin{equation}
X_t \approx A R_t A^\top, \quad t = 1, 2, \ldots, T
\end{equation}
where $A \in \mathbb{R}^{N \times R}$ represents the latent feature matrix of nodes, $R_t \in \mathbb{R}^{R \times R}$ denotes the relation matrix at time $t$, and $R$ is the decomposition rank.

\subsubsection{MLP Community Mapping}
In time slice t, an MLP mapping function $f_\theta: \mathbb{R}^R \rightarrow \mathbb{R}^K$ with parameters $\theta$ is introduced:
\begin{equation}
\mathbf{B}_t = f_\theta(A R_t)
\end{equation}

The MLP structure contains two fully connected layers with ReLU activation:
\begin{align}
h &= \text{ReLU}(W_1 \cdot (A R_t) + b_1) \\
\mathbf{B}_t &= \text{Softmax}(W_2 h + b_2)
\end{align}
where $\theta = \{W_1, b_1, W_2, b_2\}$ represents the MLP parameters.

\subsubsection{Modularity Maximization Refinement}
The improved Louvain algorithm refines the initial community partition,and
the modularity of a dynamic network at time $t$ is defined as:
\begin{equation}
Q(t) = \frac{1}{2L^{(t)}} \sum_{ij} \left[ w_{ij}^{(t)} - \frac{d_i^{(t)} d_j^{(t)}}{2L^{(t)}} \right] \delta(c_i^{(t)}, c_j^{(t)})
\end{equation}
where $L^{(t)}$ denotes the total number of edges at time $t$, $d_i^{(t)}$ represents the degree of node $i$ at time $t$, and $\delta$ is the indicator function\cite{6748996,7900340,luo}.

\subsection{Loss Function}
The model's loss function consists of reconstruction loss and regularization terms:
\begin{equation}
\mathcal{L} = \mathcal{L}_{\text{rescal}} + \mathcal{L}_{\text{mlp}}
\end{equation}

\textbf{RESCAL decomposition loss}:
\begin{equation}
\mathcal{L}_{\text{rescal}} = \frac{1}{2} \sum_{t=1}^T \| X_t - A R_t A^\top \|_F^2 + \frac{\lambda_A}{2} \| A \|_F^2 + \frac{\lambda_R}{2} \sum_{t=1}^T \| R_t \|_F^2
\end{equation}

\textbf{Community reconstruction loss}:
\begin{equation}
\mathcal{L}_{\text{mlp}} = \sum_{t=1}^T \| X_t - \tilde{B}_t (\tilde{B}_t)^\top \|_F^2+\beta \cdot\sum_{t=2}^{T} \left\| \mathbf{B}_{(t)} - \mathbf{B}_{(t-1)} \right\|_F^2
\end{equation}
where $X_t$ represents the adjacency matrix of the dynamic network at time $t$ and $\tilde{B}^t$ denotes the normalized community indicator matrix.The second term of MLP loss penalizes abrupt changes in community structures between consecutive time slices, ensuring coherent temporal evolution patterns\cite{wang2024mlp}.

\subsection{Optimization Algorithm}
An alternating optimization strategy is employed:
\begin{enumerate}
\item Update $A$ and $R_t$ in RESCAL decomposition using multiplicative update rules:

\begin{equation}
A \leftarrow A \cdot \frac{\sum_{t} X_t A R_t^\top + X_t^\top A R_t}{\sum_{t} (R_t A^\top A R_t^\top + R_t^\top A^\top A R_t) + \lambda_A I}
\end{equation}

\begin{equation}
R_t \leftarrow R_t \cdot \frac{A^\top X_t A}{A^\top A R_t A^\top A + \lambda_R R_t}
\end{equation}

\item Fix RESCAL parameters and update MLP parameters $\theta$ via gradient descent
\item Obtain initial community division and input it into the modularity maximization algorithm for further optimization
\end{enumerate}

The specific steps of the model are summarized in Algorithm 1.

\begin{algorithm}
\caption{MLP-NTD for Dynamic Community Detection}
\label{alg:mlp-NTD}
\begin{algorithmic}[1]
\State \textbf{Input:} 
\State \quad Dynamic network $\mathbf{G}$ \quad Parameters $\lambda_A$, $\lambda_R$
\State \quad Maximum number of iterations $\varphi$
\State \quad Rank $R$ \quad Number of communities $k$
\State \textbf{Output:}
\State \quad List of community members for each time segment $\mathbf{C} = (c_1, c_2, \ldots, c_T)$
\State
\State Initialize number of iterations $n = 0$, list $\mathbf{C} = \emptyset$
\State Randomly initialize matrix $A$ and tensor $\mathbf{R}$
\While{$n < \varphi$ \textbf{and} not converged}
    \State Update $A$ using Equation (10)
    \State Update $\mathbf{R}$ using Equation (11)
    \State $n = n + 1$
\EndWhile
\State Obtain initial community structure $\mathbf{C}$ using Equations (2) and (4)
\State Refine community structure $\mathbf{C}$ using modularity maximization algorithm
\end{algorithmic}
\end{algorithm}
\subsection{Algorithm Analysis}
\subsubsection{Computational Complexity}
The computational complexity of the MLP-NTD framework is analyzed by examining its major components: RESCAL tensor decomposition, MLP mapping, and modularity optimization.

The computational complexity of MLP-NTD is: $\mathcal{O}(T N^2 R)$, which is comparable to standard tensor decomposition methods while providing enhanced community detection capabilities through the MLP enhancement.

\subsubsection{Memory Complexity}
The memory requirements are primarily determined by storing the adjacency tensor and intermediate representations:$ \mathcal{O}(N^2 T)
\label{eq:memory_complexity}$

This demonstrates that MLP-NTD maintains reasonable memory efficiency while incorporating the neural network enhancement for improved community detection.
\section{Experiments}
\label{sec:experiments}

This section presents a comprehensive experimental evaluation of the proposed MLP-NTD framework. The experimental setup encompasses dynamic network datasets, baseline comparison methods, and performance metrics, followed by detailed analysis of the obtained results.

\subsection{Datasets and Comparison Methods}
\label{subsec:datasets_methods}

\subsubsection{Datasets Description}
Two real-world dynamic network datasets were employed to evaluate the performance of the proposed method:

\begin{itemize}
    \item \textbf{Chess}: A temporal network recording chess games among players from 1998 to 2006, where edges represent games between players and temporal slices correspond to yearly intervals.
    
    \item \textbf{Cellphone}: A mobile communication network capturing call records among individuals over a 10-day period, with edges representing call interactions and temporal slices organized by daily intervals.
\end{itemize}

Table~\ref{tab:dataset_statistics} provides detailed statistical characteristics of the experimental datasets, including network scale, temporal coverage, and structural properties.

\begin{table}[htbp]
\centering
\caption{Statistical Characteristics of Dynamic Network Datasets}
\label{tab:dataset_statistics}
\begin{tabular}{lcll}
\toprule
Dataset &Nodes   & edges&Time slices\\
\midrule
Chess &7,301   & 66833&9\\
Cellphone &400   & 512&10\\
\bottomrule
\end{tabular}
\end{table}

\subsubsection{Comparison Methods}
Five state-of-the-art dynamic community detection methods were selected for comparative analysis:

\begin{itemize}
    \item \textbf{TMOGA}~\cite{zou2021transfer}: A multi-objective genetic algorithm approach incorporating transfer learning for temporal network analysis
    
    \item \textbf{Cr-ENMF}~\cite{ma2020co}: A co-regularized nonnegative matrix factorization framework with temporal smoothing constraints
    
    \item \textbf{DECS}~\cite{liu2020detecting}: An evolutionary community detection method based on genetic algorithms for dynamic social networks
    
    \item \textbf{MENCPD}: A modularity-enhanced nonnegative tensor CP decomposition approach serving as an ablation baseline
    
    \item \textbf{MLP-NTD (proposed)}: Our proposed MLP-enhanced nonnegative tensor RESCAL decomposition method with rank-community number decoupling
\end{itemize}

\subsubsection{Evaluation Methodology}
The effectiveness of all methods was evaluated using \textbf{Modularity} as the primary performance metric. Modularity quantifies the quality of community partitions by measuring the difference between actual edge densities within communities and expected densities in random network configurations. Higher modularity values indicate better community structures with denser intra-community connections and sparser inter-community connections.

\subsection{Experimental Results and Analysis}
\label{sec:results}

\subsubsection{Modularity Performance Comparison}
\begin{figure}[htbp]
\centering
\begin{minipage}{0.48\textwidth}
\centering
\includegraphics[width=\linewidth]{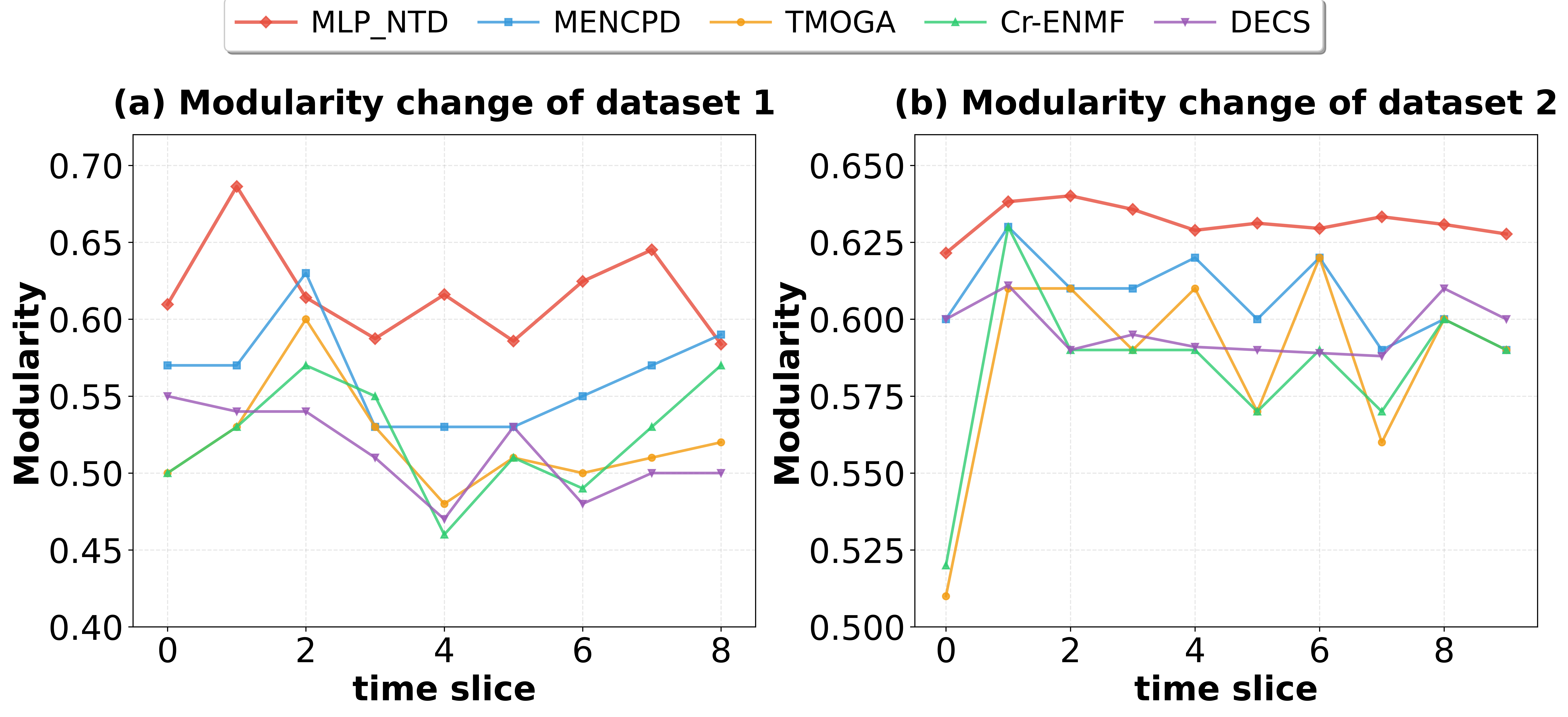}
\caption{Modularity comparison across time slices: (a) Dataset 1 (Chess), (b) Dataset 2 (Cellphone). MLP-NTD consistently achieves higher modularity values compared to baseline methods.}
\label{fig:modularity_comparison}
\end{minipage}
\hfill

\end{figure}

To evaluate the effectiveness of the proposed MLP-NTD framework, comprehensive experiments were conducted on two real dynamic networks. Figure~\ref{fig:modularity_comparison} presents modularity changes across different time slices, demonstrating the superior performance of the proposed method.

As shown in Figure~\ref{fig:modularity_comparison}, MLP-NTD maintains consistently higher modularity values across all time slices for both datasets. On Dataset 1 (Chess network), the proposed method achieves modularity values ranging from 0.55 to 0.70, significantly outperforming all baseline methods. Similarly, on Dataset 2 (Cellphone network), MLP-NTD demonstrates stable performance with modularity values between 0.58 and 0.66, showing remarkable consistency throughout temporal evolution.

\begin{figure}[htbp]
\centering
\begin{minipage}{0.48\textwidth}
\centering
\includegraphics[width=\linewidth]{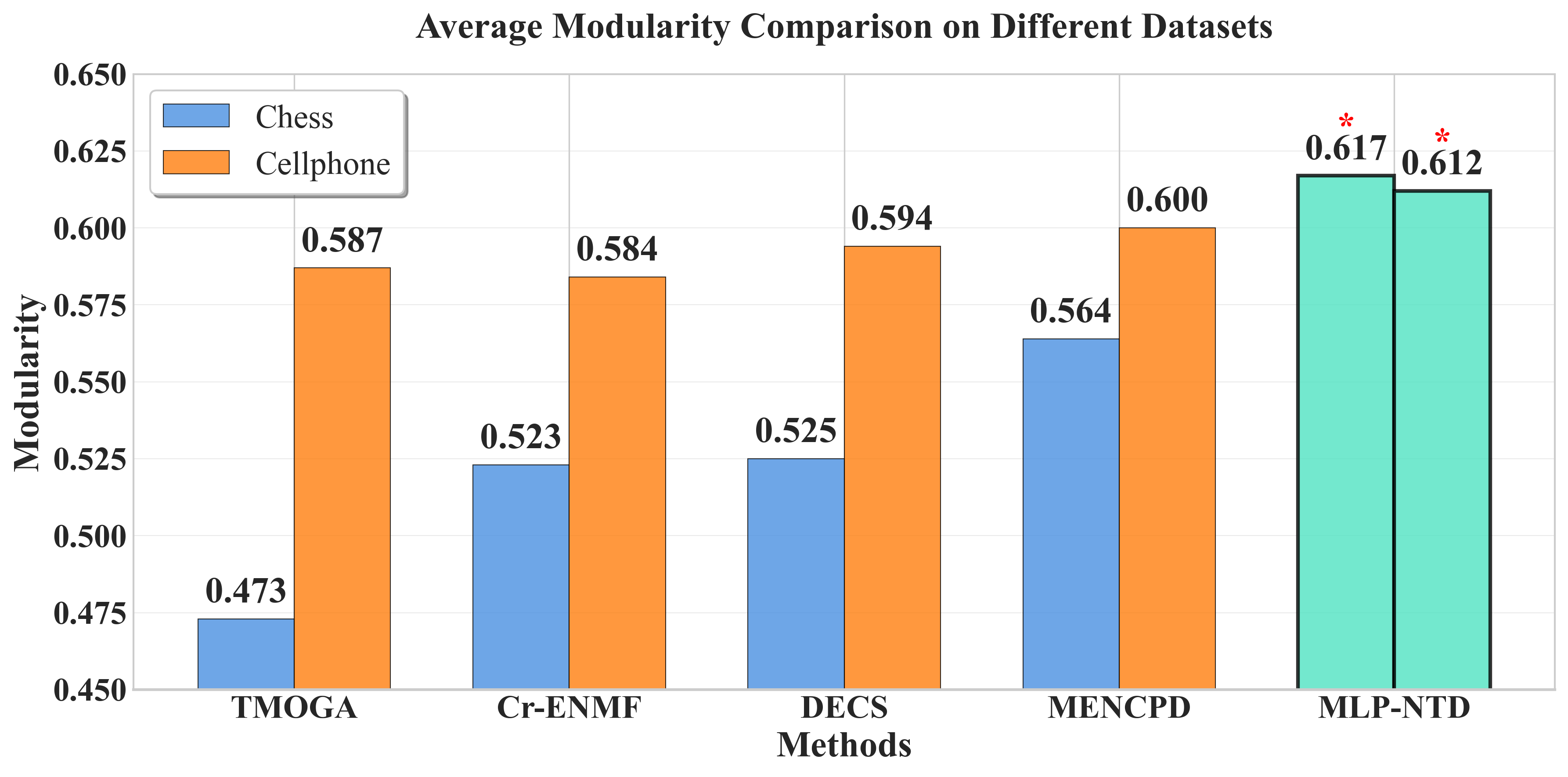}
\caption{Average Modularity Comparison across Different Dataset.}
\label{fig:average_modularity_comparison}
\end{minipage}
\end{figure}

The quantitative results are further summarized in Fig~\ref{fig:average_modularity_comparison}, where MLP-NTD achieves the highest average modularity scores across all datasets. Specifically, the proposed method attains 0.617 on the Chess dataset and 0.612 on the Cellphone dataset, representing improvements of 3.8\%-4.4\% over the original MNTD framework.
\hfill

\subsubsection{Community Structure Visualization Analysis}
\begin{figure}
\begin{minipage}{0.96\textwidth}
\centering
\begin{subfigure}{0.45\linewidth}
\centering
\includegraphics[width=\linewidth]{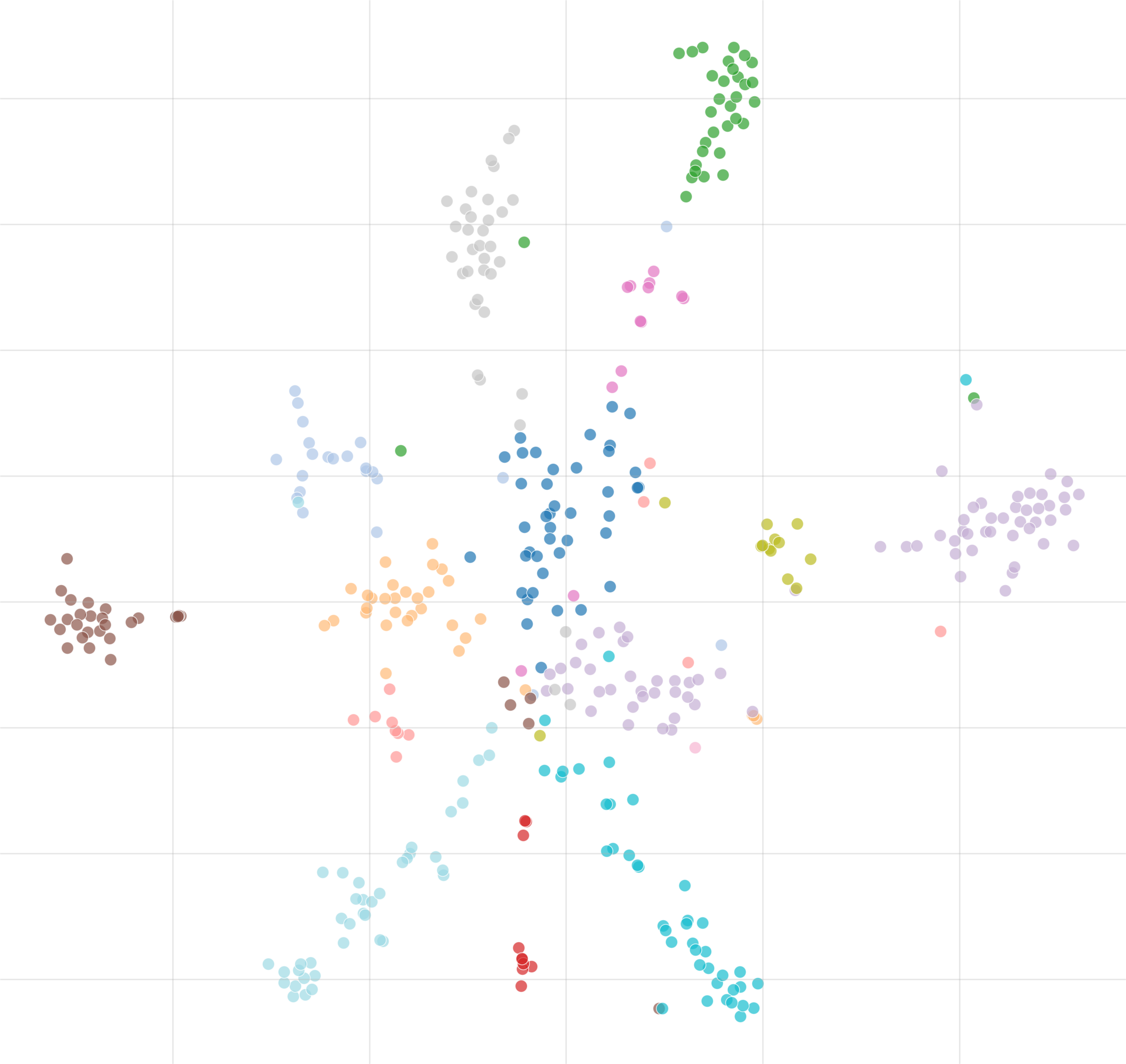}
\caption{MLP-NTD}
\label{fig:tsne_mlp}
\end{subfigure}
\hfill
\begin{subfigure}{0.45\linewidth}
\centering
\includegraphics[width=\linewidth]{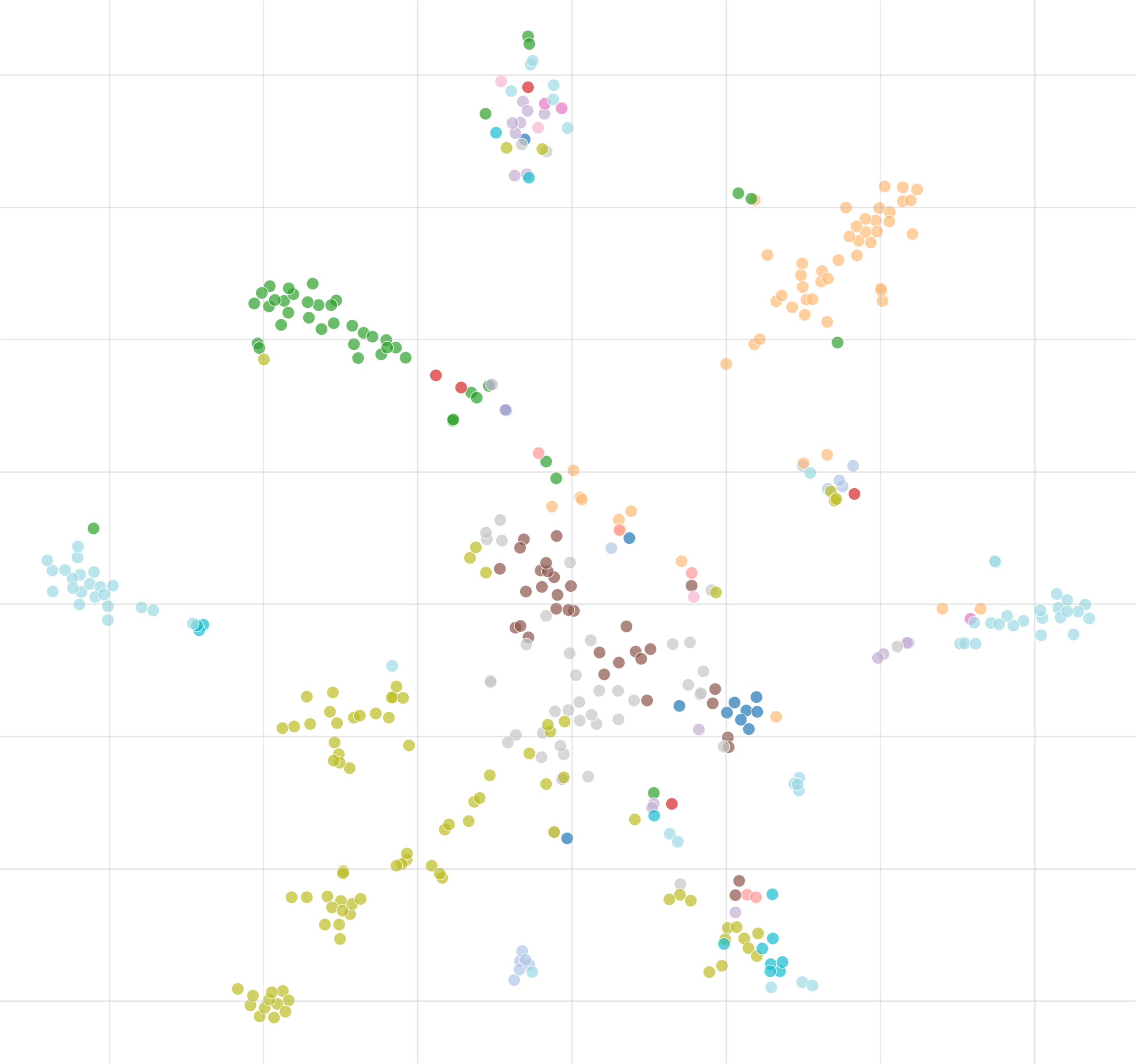}
\caption{Original MNTD}
\label{fig:tsne_NTD}
\end{subfigure}
\caption{t-SNE visualization of community structure at time slice 4. MLP-NTD exhibits clearer community separation and more compact clustering compared to the original MNTD.}
\label{fig:tsne_comparison}
\end{minipage}
\end{figure}

The t-SNE visualizations in Figure~\ref{fig:tsne_comparison} provide compelling evidence for the improved community detection capability of MLP-NTD. At time slice 4, the proposed method (Figure~\ref{fig:tsne_mlp}) demonstrates significantly clearer community boundaries and more compact clustering compared to the original MNTD (Figure~\ref{fig:tsne_NTD}).

Key observations from the visualization include:

\begin{itemize}
\item \textbf{Enhanced Community Separation}: MLP-NTD produces well-separated clusters with minimal overlap, indicating distinct community structures. Substantially larger inter-cluster distances suggest better discrimination between different communities.

\item \textbf{Improved Intra-cluster Cohesion}: Nodes within the same community are more tightly grouped in MLP-NTD, demonstrating stronger internal connections and coherence within detected communities.

\item \textbf{Reduced Noise and Outliers}: The proposed method effectively minimizes scattered nodes between clusters, resulting in cleaner community boundaries and reduced ambiguity in community assignments.

\end{itemize}

\section{Discussion and Conclusion}

\subsection{Analysis of MLP-NTD Performance}

The experimental results presented in this study consistently demonstrate the superior performance of the proposed MLP-NTD framework across multiple evaluation metrics and real-world dynamic network datasets. The proposed model introduces a multilayer perceptron-based mapping mechanism that effectively decouples the number of communities from the tensor decomposition rank, representing a fundamental advancement beyond conventional approaches that require strict equality between these parameters. This architectural innovation enables the model to overcome the inherent limitation where decomposition rank must precisely match actual community size, thereby addressing the critical challenge of severe mismatch between tensor rank and genuine community structure dimensions. Through this decoupling framework, the model achieves more accurate initial community assignments by adaptively learning optimal mappings from latent feature spaces to community partitions without being constrained by predetermined rank specifications. Experimental validation demonstrates that this approach not only resolves the performance bottleneck caused by rank-community size misalignment but also enhances the model's adaptability to diverse network structures, providing a more robust and flexible solution for dynamic community detection while maintaining computational efficiency and interpretability.

\subsection{ Future Research Directions}

We consider incorporating graph neural networks (GNNs) represents a promising direction for enhancing node representation capability. GNNs can explicitly model graph structural information through neighborhood aggregation, potentially capturing local network properties that complement the global patterns identified through tensor decomposition. This hybrid approach could leverage both the structural awareness of GNNs and the temporal modeling strength of tensor methods.

\bibliographystyle{IEEEtran}
\bibliography{references}

\end{document}